\documentclass{article}
\usepackage{spconf,amsmath,graphicx, cite}

\usepackage{setspace}
\usepackage{epstopdf}
\usepackage{times}
\usepackage[table]{xcolor}
\usepackage[draft]{hyperref}
\usepackage{comment}
\usepackage{fancyvrb}
\usepackage{listings}
\usepackage{rotating}
\usepackage{setspace}
\usepackage{subcaption}
\usepackage{lipsum}
\usepackage{booktabs, multicol, multirow}
\usepackage[ruled,vlined,linesnumbered]{algorithm2e}
\usepackage{algorithmic}

\begin{document}
\title{QuantPipe: Applying Adaptive Post-Training Quantization for Distributed Transformer Pipelines in Dynamic Edge Environments}
%
\name{Haonan Wang$^{1}$ $^{2}$, Connor Imes$^{2}$, Souvik Kundu$^{1}$ $^{3}$, Peter A. Beerel$^{1}$ $^{2}$, Stephen P. Crago$^{1}$ $^{2}$, John Paul Walters$^{2}$
\address{
     $^{1}$University of Southern California, Los Angeles, USA\\
     $^{2}$USC Information Sciences Institute, Los Angeles, USA\\
     $^{3}$Intel Labs, San Diego, USA
    }
}
%
%
%
%
\maketitle
\begin{abstract}
Pipeline parallelism has achieved great success in deploying large-scale transformer models in cloud environments, but has received less attention in edge environments. Unlike in cloud scenarios with high-speed and stable network interconnects, dynamic bandwidth in edge systems can degrade distributed pipeline performance.
We address this issue with QuantPipe, a communication-efficient distributed edge system that introduces post-training quantization (PTQ) to compress the communicated tensors.
QuantPipe uses adaptive PTQ to change bitwidths in response to bandwidth dynamics, maintaining transformer pipeline performance while incurring limited inference accuracy loss. We further improve the accuracy with a directed-search analytical clipping for integer quantization method (DS-ACIQ), which bridges the gap between estimated and real data distributions.
Experimental results show that QuantPipe adapts to dynamic bandwidth to maintain pipeline performance while achieving a practical model accuracy using a wide range of quantization bitwidths, e.g., improving accuracy under 2-bit quantization by 15.85\% on ImageNet compared to naive quantization.

\end{abstract}
\section{Introduction}
\label{sec:intro}
Recently, transformer models~\cite{vaswani2017attention, dosovitskiy2020image, kenton2019bert} have achieved high accuracy in many natural language processing~\cite{stephen2017pointer} and computer vision tasks~\cite{deng2009imagenet, wang2020temporal}. However, this high accuracy come at the cost of extremely large size. For example, Megatron-LM has 8.3 billion parameters~\cite{shoeybi2019megatron} and GPT-3 has 175 billion parameters~\cite{brown2020language}.
To alleviate this issue, many distributed parallelism training strategies have been proposed to use decentralized cloud compute resources and accelerate the training process. For instance, the Parameter Server~\cite{li2014communication} approach is a typical data-parallel method that trains multiple duplicated models with different data sets in parallel. Another model-parallel method~\cite{lepikhin2020gshard} slices a large-scale model into several sub-models that are small enough to fit on single nodes. The data that flows across the slicing boundary now must be transmitted through the communication interface between devices. A special case of model parallelism is pipeline parallelism, which partitions the model into consecutive shards and pipelines the execution of these shards across devices. Pipeline parallelism is straightforward, but efficient, and thus widely adopted by many applications~\cite{huang2019gpipe, narayanan2019pipedream, he2021pipetransformer}.

\begin{figure}[tb]
    \centering
    \includegraphics[width=0.8\linewidth]{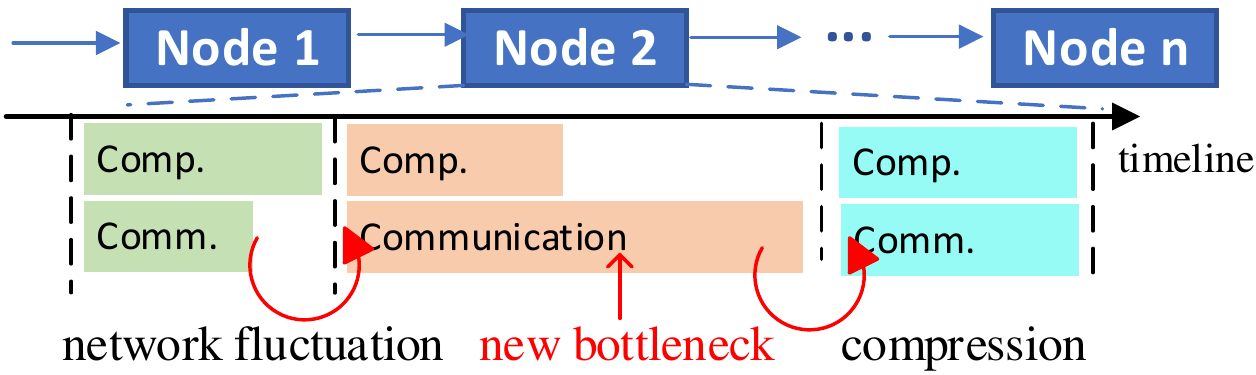}
    \caption{Performance analysis in a pipeline system.}
    \label{fig:comm}
    \vspace{-0.2in}
\end{figure}

Although pipeline parallelism has made considerable progress in training large models on the cloud, few of works focus on empowering inference at the edge. It is vital for many applications, since they may not be able to leverage the computation power on the cloud due to constraints on latency, privacy, or an unreliable (or non-existent) link to the cloud~\cite{zhang2021federated}. For example, to enable a real-time detection task on a drone formation, it may be more reliable to process the task using processors of multiple drones than offloading the task to the cloud with long round-trip latency. Distributed edge inference systems have therefore been proposed~\cite{PipeEdge,zhang2021federated}. 
However, when high-speed and stable bandwidths are not guaranteed, these systems may fail to achieve the expected performance. Fig.~\ref{fig:comm} demonstrates how the overall throughput suffers when communication bandwidth between individual stages is reduced. System performance can no longer be improved solely by further refinements to the partition strategy --- communication between nodes must be optimized.

\begin{figure*}[t]
    \centering
    \includegraphics[width=0.9\linewidth]{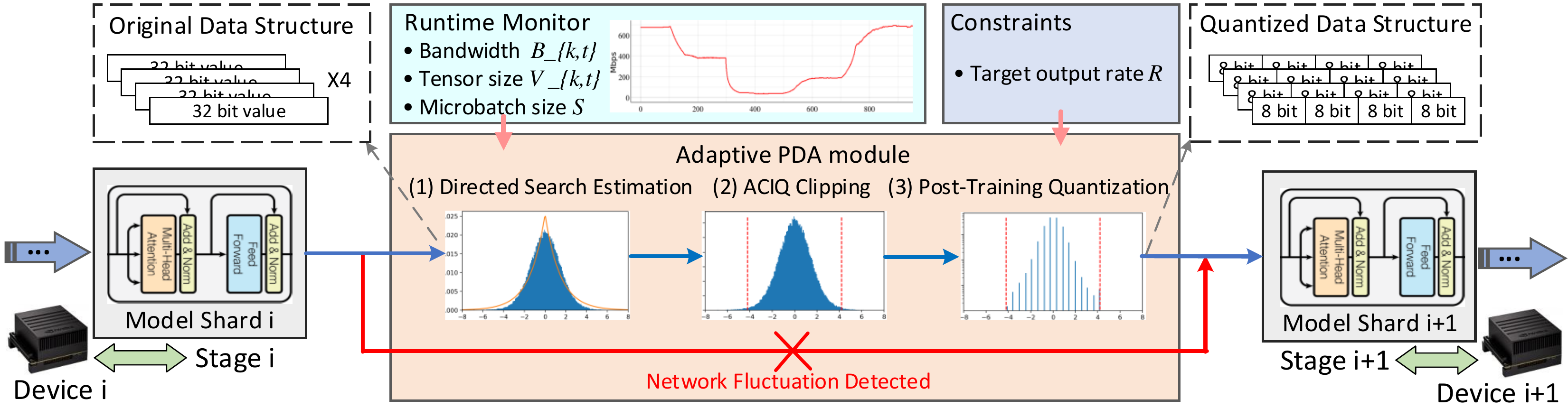}
    \caption{Overview of QuantPipe system. Adaptive PDA module is introduced for communication compression.}
    \label{fig:system}
\end{figure*}

We address this need with QuantPipe, a communication-efficient distributed edge pipeline system using post-training quantization (PTQ). 
To the best of our knowledge, we are the first work that leverages PTQ in a distributed pipeline system specifically to compress the communication of activations between pipeline stages. QuantPipe prevents bottlenecks caused by stages suffering from performance degradation resulting from intermittent bandwidth reductions in dynamic edge environments. We leverage the analytical clipping for integer quantization (ACIQ) method~\cite{banner2019post} to improve inference accuracy over a naive PTQ approach. However, we find that ACIQ still results in considerable accuracy loss under 2-bit PTQ. We find that ACIQ fails to identify the gap between the estimated and real distributions. Thus, we propose a directed-search ACIQ method (DS-ACIQ) to bridge this gap by searching for a better estimation closer to the real data distribution. Our PTQ with DS-ACIQ method (PDA) yields an accuracy improvement of 15.85\% on ImageNet, and for the first time puts 2-bit PTQ into practical implementation.
We also design an adaptive PDA module in responds to network dynamics. When bandwidth fluctuation is detected, QuantPipe applies PDA using the highest quantization bitwidth that can achieve the target performance under the current bandwidth.

\section{Background and Motivation}
\label{sec:background}

Pipeline parallelism is a straightforward but efficient parallel computing paradigm that partitions a computing task into subtasks and pipelines
their execution across devices. Several works~\cite{huang2019gpipe, narayanan2019pipedream, he2021pipetransformer} already prove it can accelerate training of large models on the cloud. However pipeline parallelism is still poorly developed for edge environments, and its downside --- the overall performance is bounded by the slowest stage --- is amplified by the intrinsic property of unstable connections in the edge environment. If any stage in the pipeline is blocked by communication due to network fluctuation, it will become the bottleneck and dominate the overall performance. 

Compressing communication in a pipeline architecture can help limit or avoid communication bottlenecks. Most existing approaches~\cite{lin2017deep, wu2022communication} focus on accelerating model training on the cloud, where connections are relatively fast and stable compared to edge environments. 
Furthermore, the pruning and knowledge distillation technique used in these solutions require training involved prior to deployment, which is too compute-intensive for edge applications.
PTQ~\cite{banner2019post} is one of the most common model compression methods that can be deployed at the runtime without any training efforts.
However, current works~\cite{banner2019post,zhao2019improving} only use PTQ to compress the model size to fit on a single device.

These challenges motivate the introduction of PTQ method to the distributed pipeline edge system. The insight is, as long as the communication becomes the new bottleneck due to network fluctuation, applying PTQ for communication compression can prevent the system from performance degradation. And we will use the popular ViT model~\cite{dosovitskiy2020image} for performance evaluation of the pipeline system, since it has a layer-wise concatenated structure without inter-layer connections, making it suitable to be partitioned by the layer boundaries in a pipeline architecture. 

\section{System Design and Implementation}
\label{sec:design}

As shown in Fig.~\ref{fig:system}, QuantPipe is developed by introducing adaptive PDA module in a pipeline system. If network fluctuation is detected at any stage, the adaptive PDA module will operate on the activation out of the previous stage, to significantly compress the communication by representing data with fewer bitwidth. For example, the communication will be compressed by 4$\times$ using 8-bit quantization. 
The adaptive PDA process consists of three steps. It first performs directed search method to get an accurate estimation of the real data distribution and then calculates the optimal ACIQ clipping range based on the estimated distribution parameter. After applying ACIQ clipping to the activation data, adaptive PDA module will conduct PTQ operation using the bitwidth that can achieve the pre-defined target output rate $R$. The required bitwidth is determined by the current measured bandwidth and some other support information collected by the runtime monitor.
The model is evenly partitioned by an optimal partition algorithm~\cite{PipeEdge}. For efficient implementation, each model shard will be assigned to only one device to avoid the overhead of collective synchronization.

\begin{figure}[tb]
    \begin{minipage}[b]{0.48\linewidth}
    \centering
    \centerline{\includegraphics[width=1.0\linewidth]{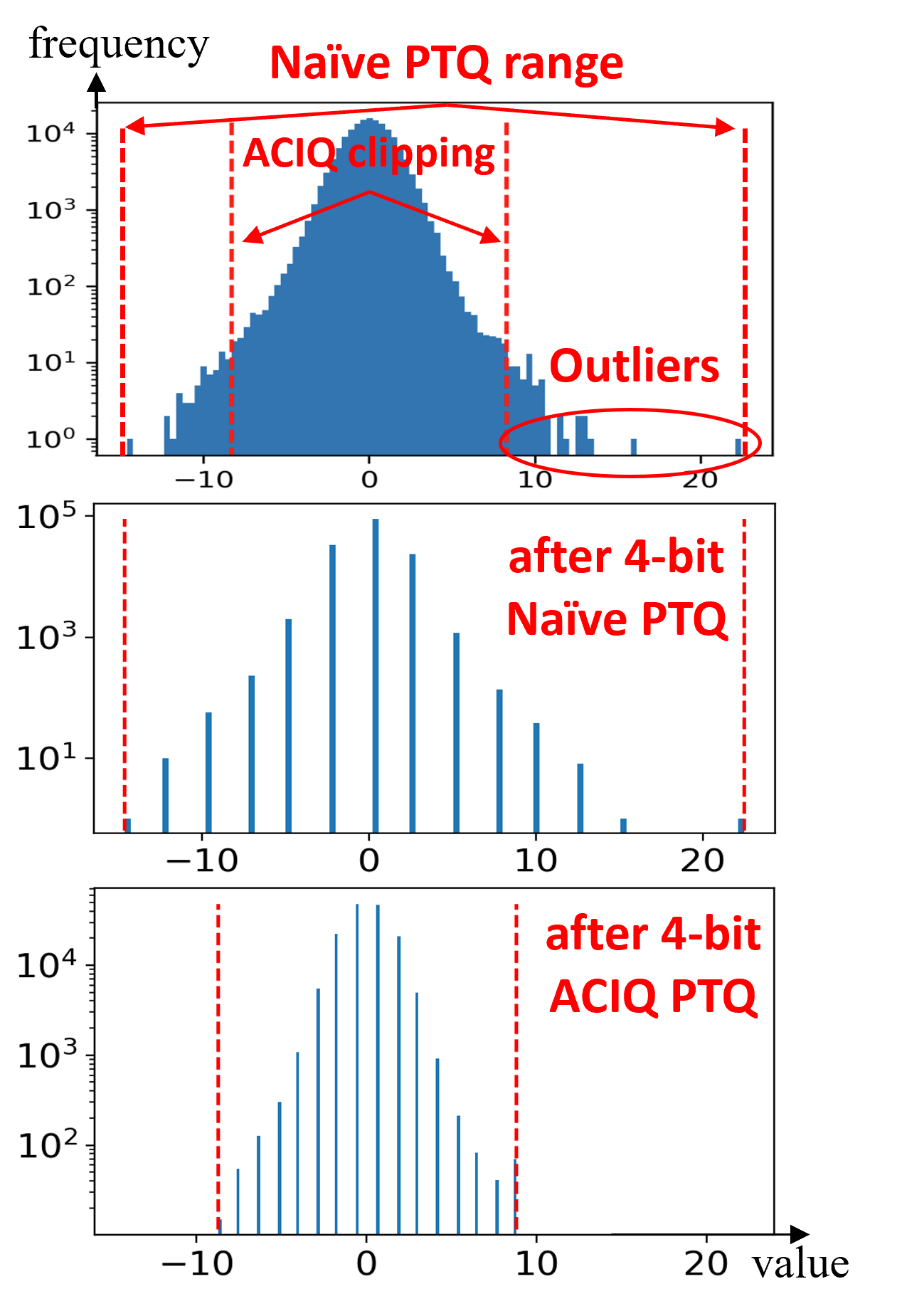}}
    \end{minipage}
    \hfill
    \begin{minipage}[b]{0.48\linewidth}
      \centering
      \centerline{\includegraphics[width=1.0\linewidth]{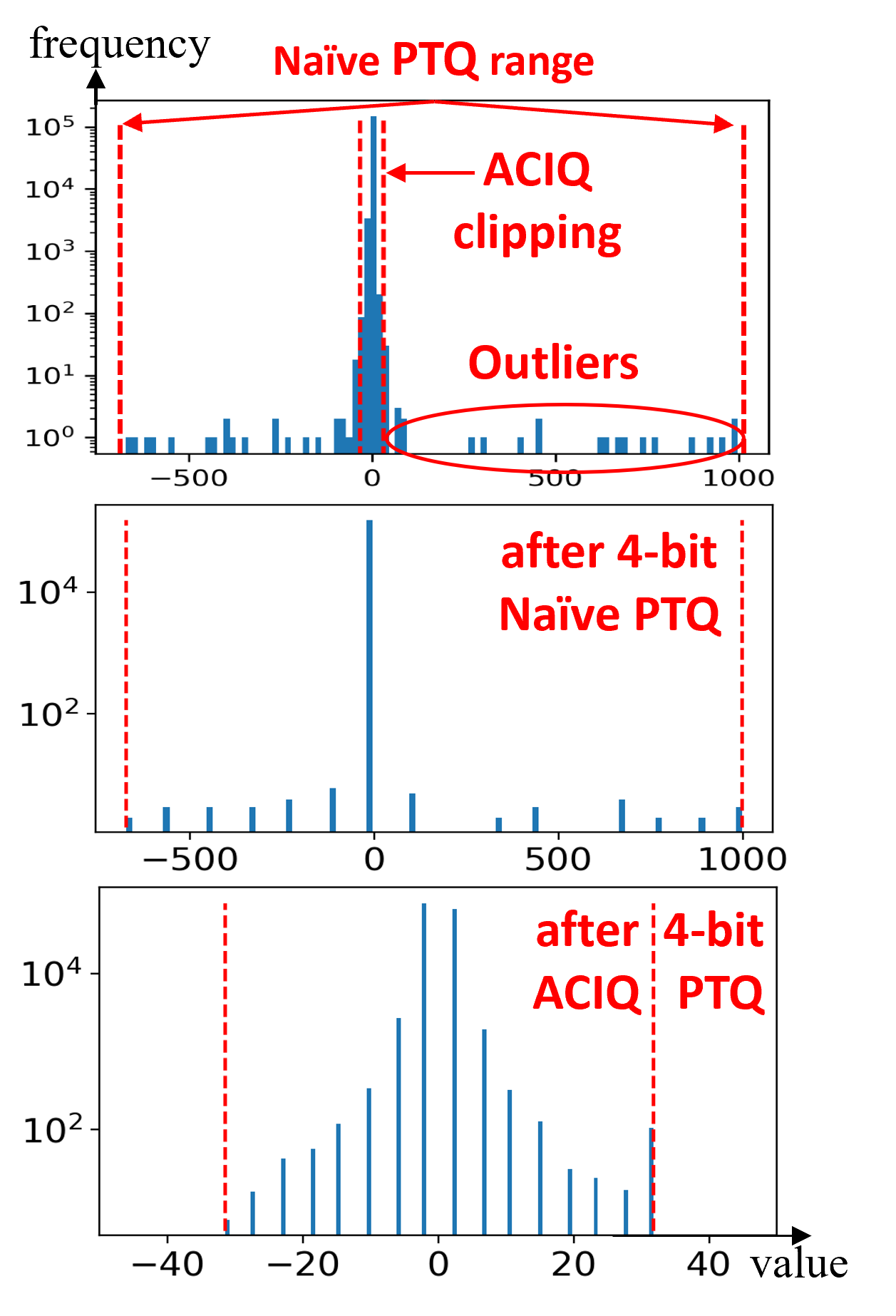}}
    \end{minipage}
    \caption{Distribution of the original data (top), after naive PTQ (middle), or after PTQ with ACIQ (bottom) from the ViT-Base model partitioned after 4th (left) and 6th (right) block.}
    \label{fig:ptq}
\end{figure}

\textbf{Naive PTQ and ACIQ.}
QuantPipe uses PTQ compression technique to achieve expected performance under limited bandwidth.
We apply uniform quantization, which divides the data range into equal intervals for rounding.
However, we find that the naive PTQ method that determines the quantization range based on the minimum and maximum tensor values can result in poor model accuracy.
As shown in Fig.~\ref{fig:ptq}, the naive quantization interval causes significant precision loss for relatively small values.
To quantify, the mean squared error (MSE) between the original tensor and the quantized tensor shows that PTQ is large and heavily influenced by outliers.
We therefore adopt the ACIQ clipping method~\cite{banner2019post} which provides a theoretical optimal clipping range that minimizes the MSE after quantization to control the downside of outliers~\cite{zhao2019improving}.
Fig.~\ref{fig:ptq} visualizes the difference between naive PTQ and ACIQ.
Precision loss is more severe for those layers with extremely large variance (e.g., at the 6th layer) where the naive PTQ derives a larger quantization interval and loses almost all information for relatively small values, resulting in worse quantization accuracy. ACIQ assumes the activation follows a Laplace distribution $L(\mu, b)$, and the optimal clipping range $\alpha$ is determined by $\alpha = F(q)\cdot b$, where the $b$ is the scale parameter of Laplace estimated from the real data, and the $F(q)$ is a lookup function based on the quantization bitwidth $q$. With ACIQ clipping, most of the outliers are clamped and the quantization range is confined such that the quantization interval is still accurate enough to represent the distribution without rounding most of the data to zero.

\begin{figure}[tb]
    \begin{minipage}[b]{0.48\linewidth}
    \centering
    \centerline{\includegraphics[width=1.0\linewidth]{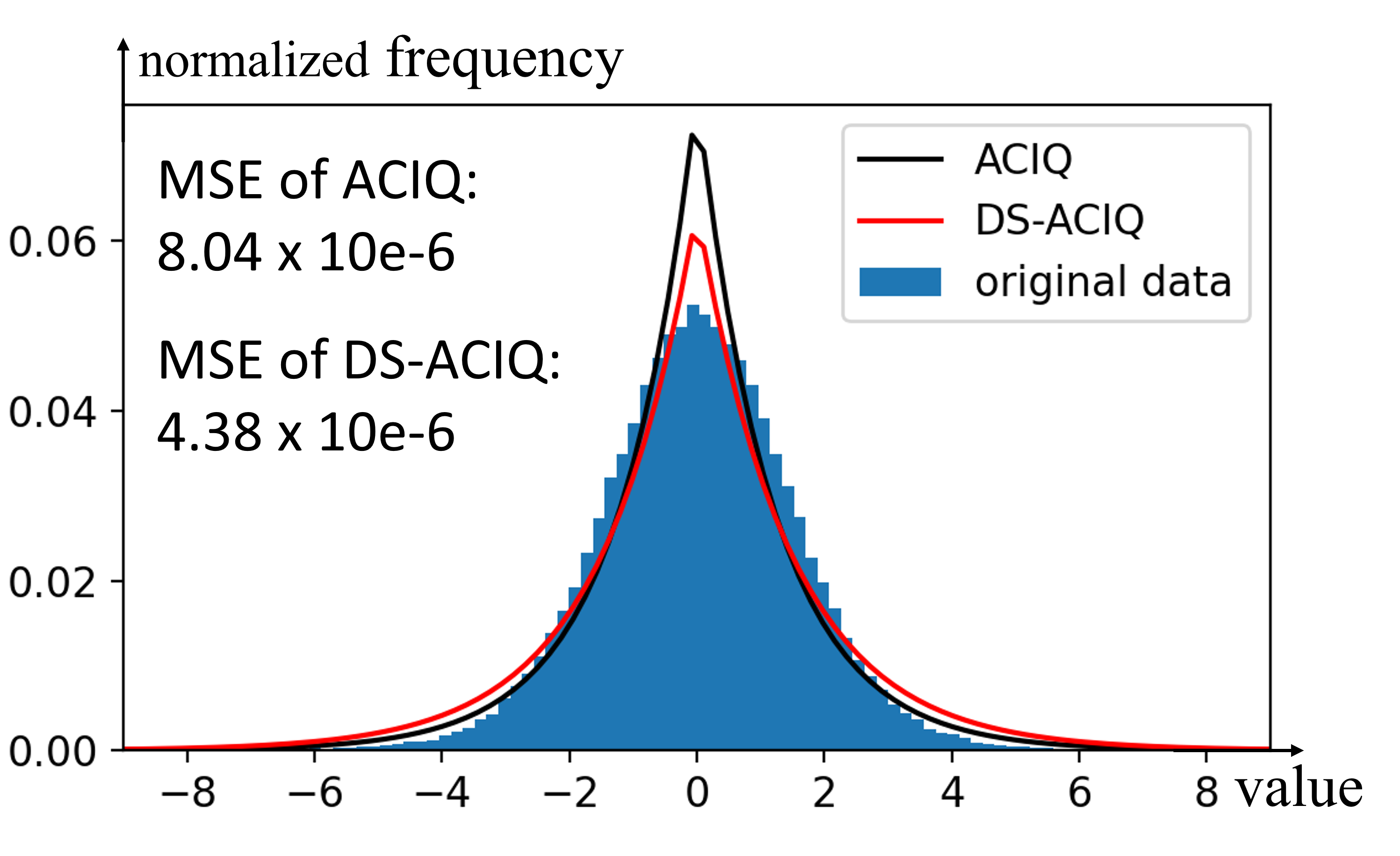}}
    \end{minipage}
    \hfill
    \begin{minipage}[b]{0.51\linewidth}
      \centering
      \centerline{\includegraphics[width=1.0\linewidth]{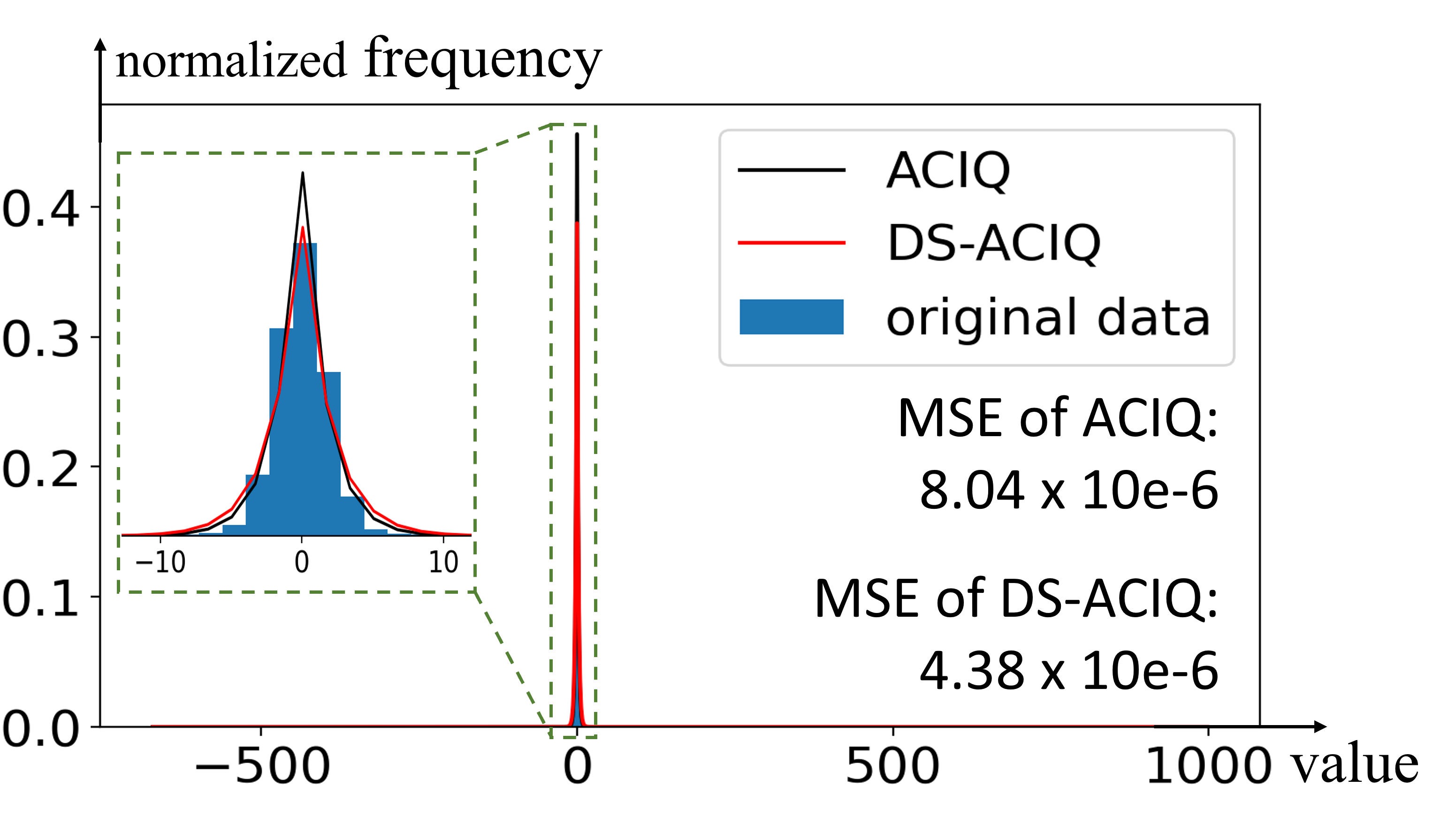}}
    \end{minipage}
    \caption{Estimated distribution by ACIQ with and without directed search after 4th (left) and 6th (right) block.}
    \label{fig:dsearch}
\end{figure}

\textbf{Directed Search.}
However, ACIQ still results in significant accuracy loss under small bitwidths, like 2-bit quantization.
Empirically, it results from the gap between the distribution estimated by ACIQ and the original distribution as shown in Fig.~\ref{fig:dsearch}.
Therefore, we propose a directed-search ACIQ (DS-ACIQ) approach to bridge this gap by searching for a better estimated scale factor $b^{*}$ that minimizes MSE along the direction towards the real data. DS-ACIQ collects the histogram information of the original data $D_R$ and compares it with the estimated distribution $D_E$, which is derived from the estimated scale factor $b_E=\sum_i \frac{abs(X)}{N}$ for tensor $X$ with size of $N$. If $max(D_R)<max(D_E)$, then it will solve the optimization problem
\begin{equation}
\operatorname*{argmin}_{b^{*}\in[b_E,b_R]} MSE(D_R,D_E) 
\end{equation}
by numerically searching $b^{*}$ in the increasing direction with $t$ steps, vice versa, where $b_R=[2\times max(D_R)]^{-1}$ is the search boundary determined by the peak of the real distribution. $t$ is heuristically set as 100 in experiments. It either finds the parameter $b^{*}$ that gives a lower MSE or otherwise use the $b_E$. It is shown in Fig.~\ref{fig:dsearch} that DS-ACIQ decreases the MSE by around 50\%. The computing overhead of DS-ACIQ averages less than 1\% in deployment.

\begin{table}[t]
\caption{Average ViT-Base model accuracy with ImageNet.
}
\label{tab:accloss}
\scalebox{0.82}{
\begin{tabular}{|l|c|c|c|c|c|c|}
\hline
         & 32bit          & 16bit   & 8bit    & 6bit     & 4bit    & 2bit    \\ \hline
PTQ      & \multirow{3}{*}{80.23\%} & \textbf{80.26}\% & 75.74\% & 43.03\% & 30.29\% & 0.44\%  \\ \cline{1-1} \cline{3-7} 
ACIQ &                          & 80.03\% & \textbf{79.35}\% & \textbf{78.87}\%  & 76.46\% & 54.97\% \\ \cline{1-1} \cline{3-7} 
PDA      &                          & 78.94\% & 78.72\% & 78.21\%  & \textbf{77.34}\% & \textbf{70.82}\% \\ \hline
\end{tabular}}
\end{table}

Table~\ref{tab:accloss} reports the average accuracy for all ViT-Base model partitions.
By leveraging DS-ACIQ, PDA limits accuracy loss to an acceptable level compared to naive PTQ.
The slight decrease of accuracy under 6,8, and 16 bit when compared with ACIQ method results from the inconsistency between the intermediate minimal-MSE representation and the final output. Although the difference is minor, DS-ACIQ approach is only activated under 4- and 2-bit quantization. For the first time, our PDA achieves a practical level of accuracy under 2 bits, outperforming that of ACIQ method by 15.85\%. 

\textbf{Adaptive PDA.}
QuantPipe's adaptive PDA module monitors the output bandwidth $B_i$ on each stage$_i,i=1\ldots n$. If a significant change in bandwidth is detected at any stage$_k$, it will estimate the bitwidth $q_{k,t+1}$ required to achieve the empirically pre-defined target sending rate $R$ under current bandwidth $B_{k,t}$ and bitwidth $q_{k,t}$ at the $t$-th iteration. It follows the equation 
Eq.~\ref{eq:adaptivePDA},
\begin{equation}\label{eq:adaptivePDA}
\begin{aligned}
    q_{k,t+1} = 32/ 2^{\lceil log(\frac{V\times 32/q_{k,t}}{S/R\times B_{k,t}})\rceil},
\end{aligned}
\end{equation}
where $V_{k,t}$ represents the size of quantized data using bitwidth of $q_{k,t}$ out from stage$_k$, and $S$ denotes the microbatch size.
Thus, QuantPipe adaptively changes the quantization bitwidth to prevent performance degradation under dynamic network conditions.


\section{Experimental Evaluation}
\label{sec:experiment}

\subsection{Experimental Setup}

Our evaluation environment is a testbed with 6 NVIDIA Jetson AGX Orin devices.
Each device has a 12-core ARM CPU, a 1792-core GPU, and runs Linux kernel 5.10.65-tegra.
We change bandwidth between devices using Linux traffic control tools (\texttt{tc}).
We implement QuantPipe on top of the PipeEdge distributed edge computing framework~\cite{PipeEdge} using Python 3.8 and PyTorch 1.12.

\subsection{Adaptive Quantization to Dynamic Bandwidth}
We demonstrate QuantPipe adapting quantization bitwidth between two nodes (stage$_1$ and stage$_2$) in response to dynamic bandwidth to maintain a send performance constraint. It is sufficient to show results on 2 nodes, since the analysis will be similar if the communication bottleneck is located at the same place in a multiple-node system.
As is common in adaptive runtime systems, QuantPipe measures relevant metrics over a window period, then makes an adaptive decision based on the window average values. We set a window period of 50 microbatches, a microbatch size of 64 images.
At roughly 200-microbatch intervals, we change the bandwidth between the devices.
Importantly, QuantPipe is not informed of this change, but rely on its own runtime measurements.

\begin{figure}[tb]
    \centering
    \includegraphics[width=1.0\linewidth]{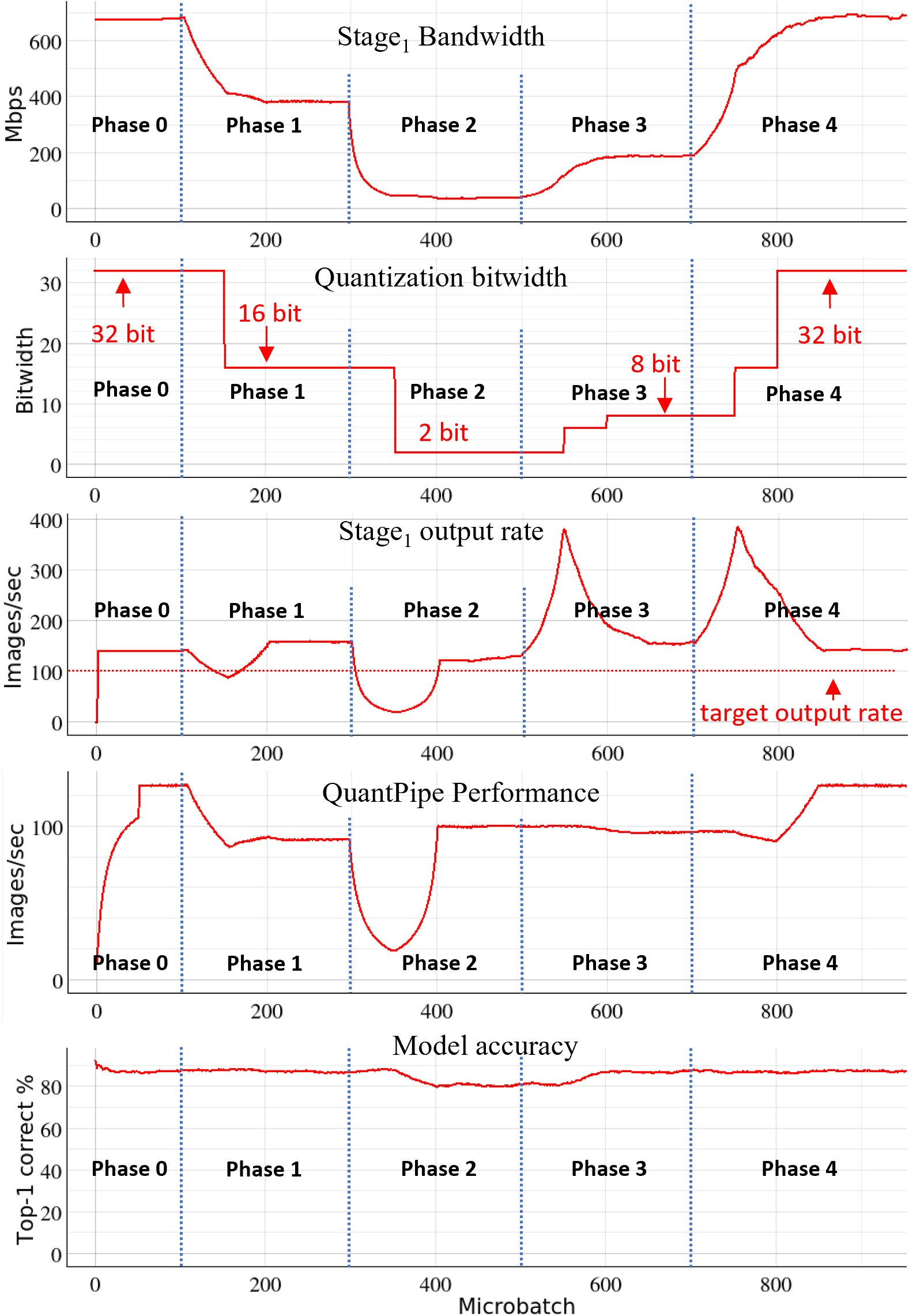}
    \caption{Evaluation of the adaptivity of QuantPipe.}
    \label{fig:adaptive}
\end{figure}

Fig.~\ref{fig:adaptive} demonstrates QuantPipe adapting to changing bandwidth in five phases.
Based on the measured bandwidth, QuantPipe's adaptive PDA module updates bitwidth after every window period.
In \textbf{Phase 0} the QuantPipe performance reaches a little higher than 100 images/sec.
At the beginning of \textbf{Phases 1 and 2}, we change the bandwidth between stage$_1$ and stage$_2$ to 400 and 50 Mbps, respectively.
On stage$_1$, QuantPipe measures that the output rate falls below the constraint value.
When the window period expires, QuantPipe applies PDA, which updates the bitwidth to 16 and then 2 bits.
The output rate of the stage$_1$ then recovers to again satisfy the target output rate, which also results in a corresponding recovery in the performance of QuantPipe. Under 2 bit, only a small reduction in pipeline accuracy is incurred.
In \textbf{Phase 3}, we increase the available bandwidth to 200 Mbps.
The output rate of stage$_1$ grows to almost 4$\times$ higher than the target output rate, but the overall system performance is not further improved since bandwidth is no longer the bottleneck.
PDA responds by switching from 2-bit to 6-bit quantization (as the bandwidth is still changing due to measurement latency), and then finally to 8 bit, which satisfies the target output rate constraint while maximizing the bitwidth and thus pipeline accuracy.
In \textbf{Phase 4}, we remove the bandwidth limitation and the system returns to its nominal state, i.e., running without quantization. 
Throughout all phases, the model accuracy curve remains at an acceptable level of accuracy loss, even under 2-bit quantization.
The experiment results illustrate that QuantPipe is adaptive and resistant to network fluctuation. 

\section{Conclusion}
\label{sec:conclu}
In this paper, we propose QuantPipe, a post-training quantization (PTQ) paradigm for communication compression in distributed transformer pipelines in dynamic edge environments. We introduce DS-ACIQ, a quantization method to improve accuracy loss caused by small-bitwidth PTQ. We empirically demonstrate adaptive quantization to recover from performance degradation caused by network fluctuations, with only limited accuracy loss. 

\section*{Acknowledgments}

{\footnotesize
This work was supported by the Department of the Navy, Office of Naval Research under, NSF, and DARPA with grant \#N00014-20-1-2143, \#1763747, and \#HR00112190120, respectively. Any opinions, findings, and conclusions or recommendations expressed in this material are those of the author(s) and do not necessarily reflect the views of the funding agencies.
This work was supported in parts by NSF and DARPA with grant numbers 1763747 and HR00112190120, respectively.
}

\clearpage
\bibliographystyle{IEEEbib}
\bibliography{refs}

\begin{thebibliography}{10}

\bibitem{vaswani2017attention}
Ashish Vaswani, Noam Shazeer, Niki Parmar, Jakob Uszkoreit, Llion Jones,
  Aidan~N Gomez, {\L}ukasz Kaiser, and Illia Polosukhin,
\newblock ``Attention is all you need,''
\newblock {\em Advances in neural information processing systems}, vol. 30,
  2017.

\bibitem{dosovitskiy2020image}
Alexey Dosovitskiy, Lucas Beyer, Alexander Kolesnikov, Dirk Weissenborn,
  Xiaohua Zhai, Thomas Unterthiner, Mostafa Dehghani, Matthias Minderer, Georg
  Heigold, Sylvain Gelly, et~al.,
\newblock ``An image is worth 16x16 words: Transformers for image recognition
  at scale,''
\newblock {\em arXiv preprint arXiv:2010.11929}, 2020.

\bibitem{kenton2019bert}
Jacob Devlin Ming-Wei~Chang Kenton and Lee~Kristina Toutanova,
\newblock ``Bert: Pre-training of deep bidirectional transformers for language
  understanding,''
\newblock in {\em Proceedings of naacL-HLT}, 2019, pp. 4171--4186.

\bibitem{stephen2017pointer}
Merity Stephen, Xiong Caiming, Bradbury James, and Richard Socher,
\newblock ``Pointer sentinel mixture models,''
\newblock in {\em 5th International Conference on Learning Representations
  (ICLR)}, 2017.

\bibitem{deng2009imagenet}
Jia Deng, Wei Dong, Richard Socher, Li-Jia Li, Kai Li, and Li~Fei-Fei,
\newblock ``Imagenet: A large-scale hierarchical image database,''
\newblock in {\em 2009 IEEE conference on computer vision and pattern
  recognition}, 2009, pp. 248--255.

\bibitem{wang2020temporal}
Haonan Wang, Yuchen Mei, Jun Lin, and Zhongfeng Wang,
\newblock ``Temporal residual feature learning for efficient 3d convolutional
  neural network on action recognition task,''
\newblock in {\em 2020 IEEE Workshop on Signal Processing Systems (SiPS)}.
  IEEE, 2020, pp. 1--6.

\bibitem{shoeybi2019megatron}
Mohammad Shoeybi, Mostofa Patwary, Raul Puri, Patrick LeGresley, Jared Casper,
  and Bryan Catanzaro,
\newblock ``Megatron-lm: Training multi-billion parameter language models using
  model parallelism,''
\newblock {\em arXiv preprint arXiv:1909.08053}, 2019.

\bibitem{brown2020language}
Tom Brown, Benjamin Mann, Nick Ryder, Melanie Subbiah, Jared~D Kaplan, Prafulla
  Dhariwal, Arvind Neelakantan, Pranav Shyam, Girish Sastry, Amanda Askell,
  et~al.,
\newblock ``Language models are few-shot learners,''
\newblock {\em Advances in neural information processing systems}, vol. 33, pp.
  1877--1901, 2020.

\bibitem{li2014communication}
Mu~Li, David~G Andersen, Alexander~J Smola, and Kai Yu,
\newblock ``Communication efficient distributed machine learning with the
  parameter server,''
\newblock {\em Advances in Neural Information Processing Systems}, vol. 27,
  2014.

\bibitem{lepikhin2020gshard}
Dmitry Lepikhin, HyoukJoong Lee, Yuanzhong Xu, Dehao Chen, Orhan Firat, Yanping
  Huang, Maxim Krikun, Noam Shazeer, and Zhifeng Chen,
\newblock ``Gshard: Scaling giant models with conditional computation and
  automatic sharding,''
\newblock {\em arXiv preprint arXiv:2006.16668}, 2020.

\bibitem{huang2019gpipe}
Yanping Huang, Youlong Cheng, Ankur Bapna, Orhan Firat, Dehao Chen, Mia Chen,
  HyoukJoong Lee, Jiquan Ngiam, Quoc~V Le, Yonghui Wu, et~al.,
\newblock ``Gpipe: Efficient training of giant neural networks using pipeline
  parallelism,''
\newblock {\em Advances in neural information processing systems}, vol. 32,
  2019.

\bibitem{narayanan2019pipedream}
Deepak Narayanan, Aaron Harlap, Amar Phanishayee, Vivek Seshadri, Nikhil~R
  Devanur, Gregory~R Ganger, Phillip~B Gibbons, and Matei Zaharia,
\newblock ``Pipedream: generalized pipeline parallelism for dnn training,''
\newblock in {\em Proceedings of the 27th ACM Symposium on Operating Systems
  Principles}, 2019, pp. 1--15.

\bibitem{he2021pipetransformer}
Chaoyang He, Shen Li, Mahdi Soltanolkotabi, and Salman Avestimehr,
\newblock ``Pipetransformer: automated elastic pipelining for distributed
  training of large-scale models,''
\newblock in {\em International Conference on Machine Learning}. PMLR, 2021,
  pp. 4150--4159.

\bibitem{zhang2021federated}
Tuo Zhang, Chaoyang He, Tianhao Ma, Lei Gao, Mark Ma, and Salman Avestimehr,
\newblock ``Federated learning for internet of things,''
\newblock in {\em Proceedings of the 19th ACM Conference on Embedded Networked
  Sensor Systems}, 2021, pp. 413--419.

\bibitem{PipeEdge}
Yang Hu, Connor Imes, Xuanang Zhao, Souvik Kundu, Peter~A. Beerel, Stephen~P.
  Crago, and John~Paul Walters,
\newblock ``Pipeedge: Pipeline parallelism for large-scale model inference on
  heterogeneous edge devices,''
\newblock in {\em 2022 25th Euromicro Conference on Digital System Design
  (DSD)}, 2022, pp. 298--307.

\bibitem{banner2019post}
Ron Banner, Yury Nahshan, and Daniel Soudry,
\newblock ``Post training 4-bit quantization of convolutional networks for
  rapid-deployment,''
\newblock {\em Advances in Neural Information Processing Systems}, vol. 32,
  2019.

\bibitem{lin2017deep}
Yujun Lin, Song Han, Huizi Mao, Yu~Wang, and William~J Dally,
\newblock ``Deep gradient compression: Reducing the communication bandwidth for
  distributed training,''
\newblock {\em arXiv preprint arXiv:1712.01887}, 2017.

\bibitem{wu2022communication}
Chuhan Wu, Fangzhao Wu, Lingjuan Lyu, Yongfeng Huang, and Xing Xie,
\newblock ``Communication-efficient federated learning via knowledge
  distillation,''
\newblock {\em Nature communications}, vol. 13, no. 1, pp. 1--8, 2022.

\bibitem{zhao2019improving}
Ritchie Zhao, Yuwei Hu, Jordan Dotzel, Chris De~Sa, and Zhiru Zhang,
\newblock ``Improving neural network quantization without retraining using
  outlier channel splitting,''
\newblock in {\em International conference on machine learning}. PMLR, 2019,
  pp. 7543--7552.

\end{thebibliography}

\end{document}